# Machine learning approach for text and document mining

Vishwanath Bijalwan[1], Pinki Kumari[2], Jordan Pascual[3] and Vijay Bhaskar Semwal[4]
[1]Asst. Prof., Institute of technology Gopeshwar, Chamoli, Uttarakhand, India
[2]Bansathali University, Rajasthan, India
[3]Department of Computer Science, University of Oviedo, Spain
[4,]Senior Software Architect, Siemens Information Systems Limited, Gurgaon, India

*Abstract*

*Text Categorization (TC), also known as Text Classification, is the task of automatically classifying a set of text documents into different categories from a predefined set. If a document belongs to exactly one of the categories, it is a single-label classification task; otherwise, it is a multi-label classification task. TC uses several tools from Information Retrieval (IR) and Machine Learning (ML) and has received much attention in the last years from both researchers in the academia and industry developers. In this paper, we first categorize the documents using KNN based machine learning approach and then return the most relevant documents.*

## 1. Introduction

Information Retrieval (IR) is the science of searching for information within relational databases, documents, text, multimedia files, and the World Wide Web. The applications of IR are diverse; they include but not limited to extraction of information from large documents, searching in digital libraries, information filtering, spam filtering, object extraction from images, automatic summarization, document classification and clustering, and web searching. The breakthrough of the Internet and web search engines have urged scientists and large firms to create very large scale retrieval systems to keep pace with the exponential growth of online data. Figure below depicts the architecture of a general IR system. The user first submits a query which is executed over the retrieval system. The latter, consults a database of document collection and returns the matching document. In general, in order to learn a classifier that is able to correctly classify unseen documents, it is necessary to train it with some pre-classified documents from each category, in such a way that the classifier is then able to generalize the model it has learned from the pre-classified documents and use that model to correctly classify the unseen documents. Figure 1 shows the overview of the document indexing and retrieval system. From experiment, KNN shows the maximum accuracy as compared to the Naive Bayes and Term-Graph. The drawback for KNN is that its time complexity is high but gives a better accuracy than others.

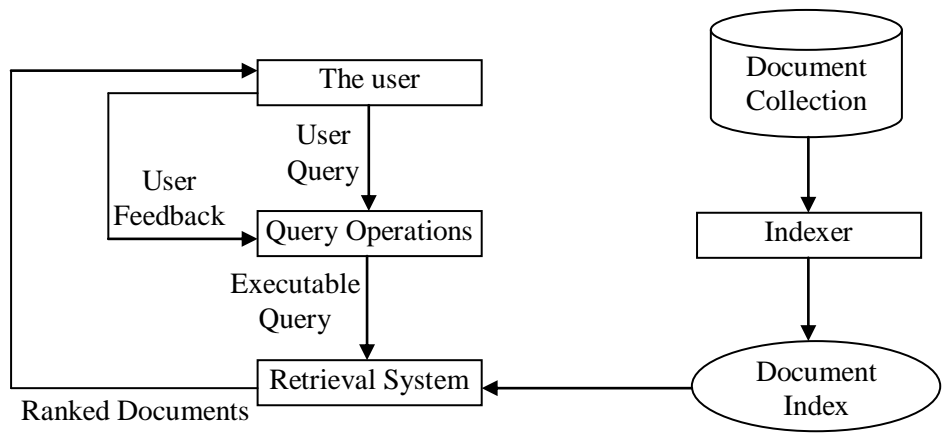

Figure 1: Overview of document retrieval system

In recent years, text categorization has become an important research topic in machine learning and information retrieval and e-mail spam filtering. It also has become an important research topic in text mining, which analyses and extracts useful information from texts. More Learning techniques have been in research for dealing with text categorization. The existing text classification methods can be classified into below six [7], [8], [9] categories:

(1) Based on Rocchio's method (Dumais, Platt, Heckerman, & Sahami, 1998; Hull, 1994; Joachims, 1998; Lam & Ho, 1998).

(2) Based on K-nearest neighbors (KNN) (Hull, 1994; Lam & Ho, 1998; Tan, 2005; Tan, 2006; Yang & Liu, 1999).

(3) Based on regression models (Yang, 1999; Yang & Liu, 1999).

(4) Based on Naıve Bayes and Bayesian nets (Dumais et al., 1998; Hull, 1994; Yang & Liu, 1999; Sahami, 1996).

(5) Based on decision trees (Fuhr & Buckley, 1991; Hull, 1994).

(6) Based on decision rules (Apte`, Damerau, & Weiss, 1994; Cohen & Singer, 1999).

Among the six types the survey aims in getting an intuitive understanding of KNN approach in which the application of various Machine Learning Techniques [11, 17, 18, 20, 21] to the text categorization problem like in the field of medicine, e-mail filtering, including rule learning for knowledge base systems has been explored. The survey is oriented towards the various probabilistic approach of KNN Machine Learning algorithm for which the text categorization aims to classify the document with optimal accuracy. Information retrieval is also used in image retrieval [19]. In recent works, to save and estimate accurate location moving object with energy constraint is proposed in [10, 14] using adaptive update algorithms. Some other recent approaches such as video summarization [12], 3D model of 2D image [13], gait pattern [15], and scale replica model [16] can also be integrated with the proposed approach to enhance the efficiency.

This paper categorizes the news articles into various categories. We work on two major scenarios:

a. Classification of documents into various categories.

Making it in the form of an application where user can upload an article and we will classify it into various categories.

- b. On entering keywords by the user we show the most relevant document for the user.

## 2. Classification Methods

This paper concerns methods for the classification of natural language text, that is, methods that, given a set of training documents with known categories and a new document, which is usually called the query, will predict the query's category.

### 2.1. Naive Bayes

The Naive Bayes classifier found its way into many applications nowadays due to its simple principle but yet powerful accuracy [2], [5]. Bayesian classifiers are based on a statistical principle. Here, the presence or absence of a word in a textual document determines the outcome of the prediction. In other words, each processed term is assigned a probability that it belongs to a certain category. This probability is calculated from the occurrences of the term in the training documents where the categories are already known. When all these probabilities are calculated, a new document can be classified according to the sum of the probabilities for each category of each term occurring within the document. However, this classifier does not take the number of occurrences into account, which is a potentially useful additional source of information. They are called "naive" because the algorithm assumes that all terms occur independent from each other. Given a set of r document vectors $D = \{d_1, \ldots, d_r\}$, classified along a set $C$ of $q$ classes, $C=\{c_1, \ldots, c_q\}$, Bayesian classifiers estimate the probabilities of each class $c_k$ given a document $d_j$ as:

$$P(c_k|d_j) = \frac{P(c_k)P(\vec{d_j}|c_k)}{P(\vec{d_j})} \tag{1}$$

In this eq. 1, $P(\vec{d_j})$ is the probability that a randomly picked document has vector $\vec{d_j}$ as its representation, and $P(c_k)$ the probability that a randomly picked document belongs to ck. Because the number of possible documents $d_j$ is very high, the estimation of $P(\vec{d_j}|c_k)$ is problematic. To simplify the estimation of $P(d_j|c_k)$, Naive Bayes assumes that the probability of a given word or term is independent of other terms that appear in the same document. While this may seem an over simplification, in fact Naive Bayes presents results that are very competitive with those obtained by more elaborate methods. Moreover, because only words and not combinations of words are used as predictors, this naive simplification allows the computation of the model of the data associated with this method to be far more efficient than other non naive Bayesian approaches. Using this simplification, it is possible to determine $P(\vec{d_j}|c_k)$ as the product of the probabilities of each term that appears in the document. So, $P(\vec{d_j}|c_k)$ may be estimated as:

$$P(\vec{d_j}|c_k) = \prod_{i=1}^{|T|} P(w_{ij}|c_k) \tag{2}$$

where, $\vec{d_j} = (w1j, \ldots, w|T|j)$.

*Algorithm:*
1) Checking the keyword in Test document and storing it in a map.
2) Calculating yes and no frequency of each keyword in the test document.
3) Calculating the probability of each keyword of the test document.

4) Classifying the Test Document into various categories on the basis of probability calculated.

**2.2. Term Graph Model**

The term graph model is an improved version of the vector space model [6] by weighting each term according to its relative "importance" with regard to term associations. Specifically, for a text document $D_i$, it is represented as a vector of term weights $D_i = <w_{1i}, ...., w_{|T|i}>$, where $T$ is the ordered set of terms that occur at least once in at least one document in the collection. Each weight $w_{ji}$ represents how much the corresponding term $t_j$ contribute to the semantics of document $d_i$. Although a number of weighting schemes have been proposed (e.g., boolean weighting, frequency weighting, tf-idf weighting, etc.), those schemes determine the weight of each term individually. As a result, important yet rich information regarding the relationships among the terms are not captured in those weighting schemes.

We introduce to determine the weight of each term in a document collection by constructing a term graph. The basic steps are as follows:

1. Preprocessing Step: For a collection of document, extract all the terms.

In our term graph model, we will capture the relationships among terms using the frequent item set mining method. To do so, we consider each text document in the training collections as a transaction in which each word is an item. However, not all words in the document are important enough to be retained in the transaction. To reduce the processing space as well as increase the accuracy of our model, the text documents need to be preprocessed by (1) remove stop words, i.e., words that appear frequently in the document but have no essential meanings; and (2) retaining only the root form of words by stemming their affixes as well as prefixes.

2. Graph Building Step:

(a) For each document, we view it as a transaction: the document ID is the corresponding transaction ID; the terms contained in the document are the items contained in the corresponding transaction. Association rule mining algorithms can thus be applied to mine the frequently co-occurring terms that occur more than minsup times in the collection.

(b) The frequent co-occurring terms are mapped to a weighted and directed graph, i.e., the term graph.

As mentioned above, we will capture the relationships among terms using the frequent item set mining method. While this idea has been explored by previous research [3], our approach distinguishes from previous approaches in that we maintain all such important associations in a graph. The graph not only reveals the important semantics of the document, but also provides a basis to extract novel features about the document, as we will show in the next section. After the preprocessing step, each document in the text collection will be stored as a transaction (list of items) in which each item (term) is represented by a unique non-negative integer. Then frequent item set mining algorithms can be used to find all the subset of items that appeared more than a threshold amount of times in the collection.

In our system, our goal is to explore the relationships among the important terms of the text in a category and try to define a strategy to make use of these relationships in the classifier and other text mining tasks. Vector space model cannot express such rich relationship among terms. Graph is thus the most suitable data structure in our context, as, in general, each term may be associated with more than one terms. We propose to use the following simple method to construct the graph from the set of frequent item sets mined from the text collections. First, we construct a node for each unique term that appears at least once in the frequent item sets. Then we create edges between two node u and v if and only if they are both contained in one frequent item set. Furthermore, we assign weights to the edges in the following way: the weight of the edge between u and v is the largest support value among all the frequent item sets that contains both of them.

For, example, consider the frequent item sets and their absolute support shown in Figure 2(a). Its corresponding graph is shown in Figure 2(b).

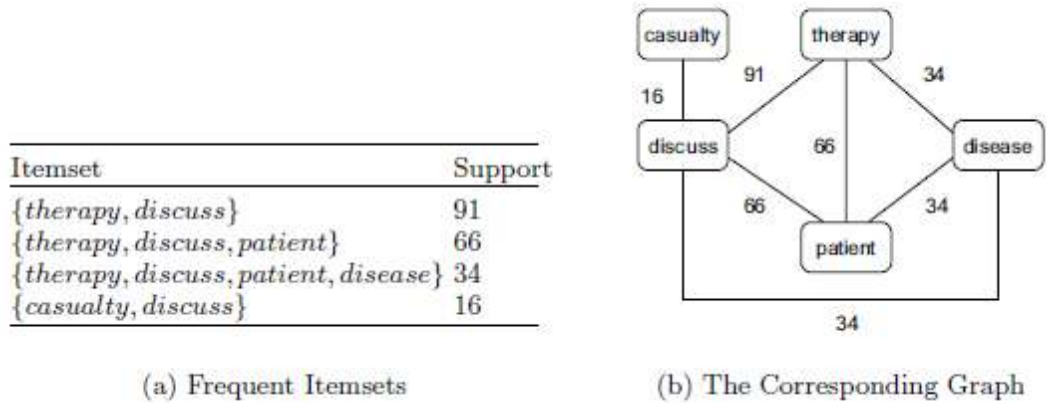

Figure 2: (a) Frequent item set with support, and (b) Corresponding graph.

*Algorithm:*
1) Setting each unique word occurring the document as nodes of the graph.
2) Making Adjacency Matrix of the keywords.
3) Making Distance Matrix using Dijkstra.
4) Calculating similarity between the test document keywords and the keywords of each category.

**2.3. k-Nearest Neighbors**

The initial application of k-Nearest Neighbors (KNN) to text categorization was reported in [4]. The basic idea is to determine the category of a given query based not only on the document that is nearest to it in the document space, but on the categories of the k documents that are nearest to it. Having this in mind, the Vector method can be viewed as an instance on the KNN method, where k=1. This work uses a vector-based, distance-weighted matching function, as did Yang, by calculating document's similarity like the Vector method.

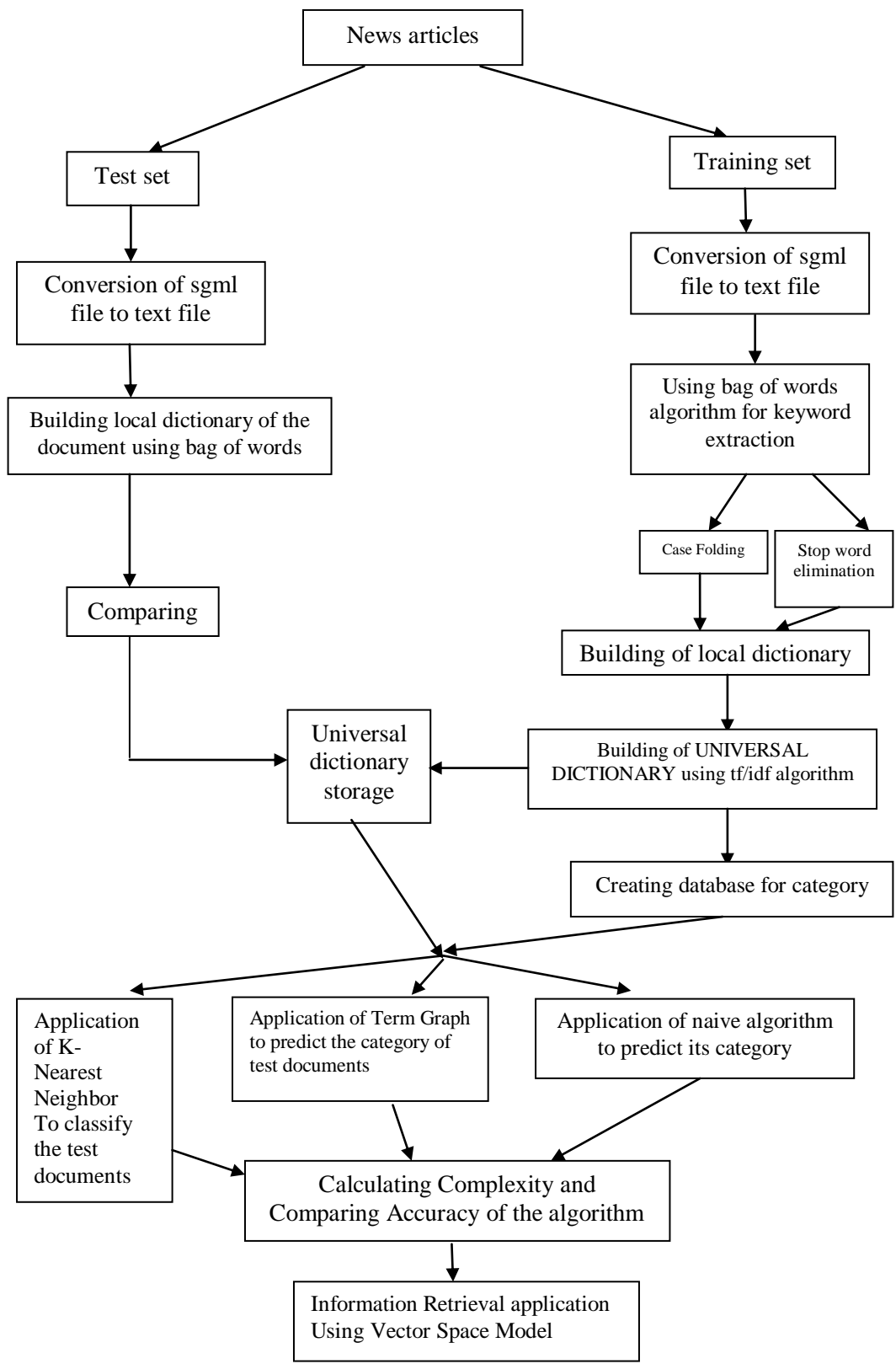

Figure 3: Flow chart of the text and document mining.

Then, it uses a voting strategy to find the query's class: each retrieved document contributes a vote for its class, weighted by its similarity to the query. The query's possible classifications will be ranked according to the votes they got in the previous step.

*Algorithm:*
1) Make vector for every document in the test set.
2) Make centriod vector for each class.
3) Calculate similarity between each document vector and class vector.
4) Document belongs to the class for which the similarity is maximum.

Figure 3 shows the flowchart of the proposed system for text and document mining using machine learning techniques.

## 3. Experimental Results

### 3.1. Dataset

The data set used for this paper is in the form of sgml files [3]. We have used Reuters-21578 dataset which is available at [1]. There are 21578 documents; according to the 'ModApte' split: 9603 training docs, 3299 test docs and 8676 unused docs. They were labeled manually by Reuters personnel. Labels belong to 5 different category classes, such as 'people', 'places', 'Exchange', 'Organization' and 'topics'. The total number of categories is 672, but many of them occur only very rarely. The dataset is divided in 22 files of 1000 documents delimited by SGML tags.

### 3.2. Implementation

For classifying the documents in Reuter-21578 we initially pre-processed the data by performing various techniques:
a. Bag of words
b. Stop word removal
c. TF-IDF
d. Case Folding
e. Normalization

Then after pre-processing, we applied KNN, Term Graph algorithm, and Naïve Bayes algorithms to classify the documents in the training set into five categories (exchange, organization, people, places and topics). We further applied our classifier model on the test documents and calculated the accuracy by comparing it with the default answers given for the test documents. To compare the above mentioned algorithms, we used the following metric:

Accuracy, which is defined as the percentage of correctly classified documents, is generally used to evaluate single-label TC tasks.

$$Accuracy = \frac{\#Correctly\ classified\ documents}{\#Total\ documents}$$

We then created an application where user can input some keywords and based on the algorithm showing higher accuracy we show the relevant document to the user.

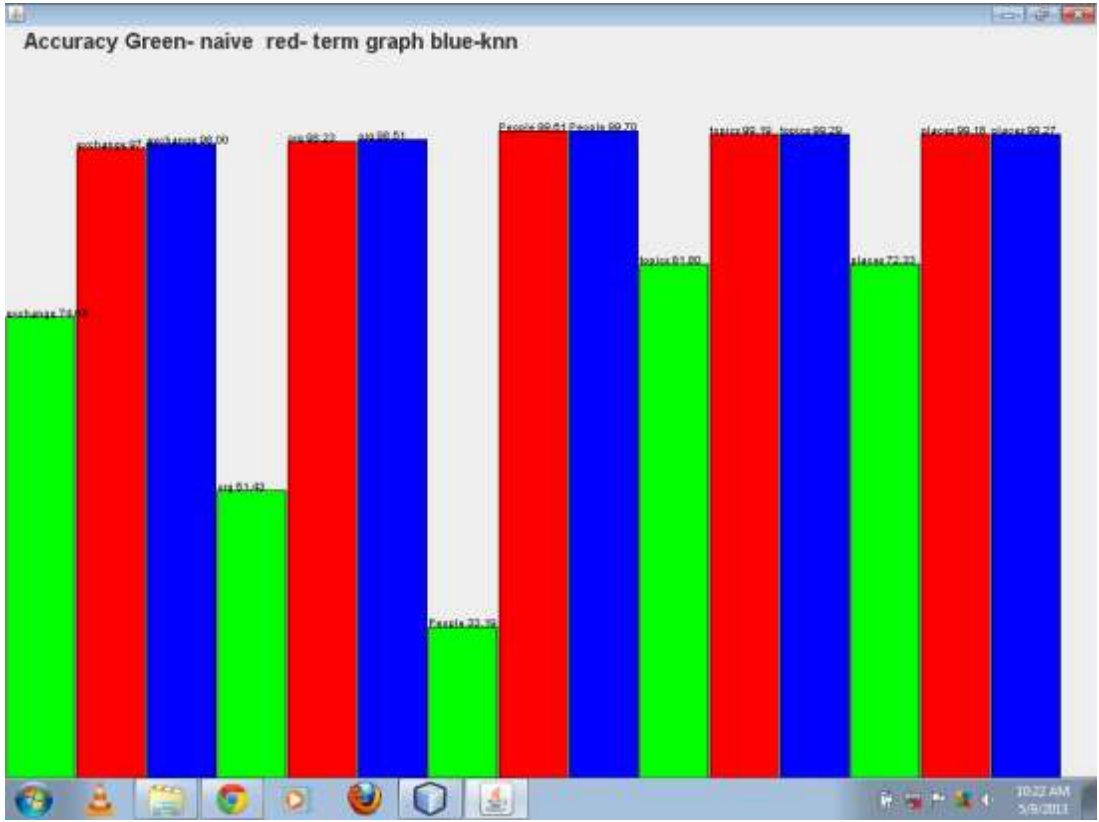

Figure 4: Graph representation of accuracy using each approach.

### 3.3. Results

We compared the accuracy of Naïve Bayes, Term Graph and KNN for Text and Document classification of our articles of Reuter 21578. As shown in the graph of Fig. 4, we found that KNN shows the best result with accuracy as provided in Table 1.

Table 1: Accuracy for each method

| Category/Method | NAÏVE | Term Graph | KNN |
|---|---|---|---|
| EXCHANGE | 74.68 | 97.41 | 98.00 |
| ORGANIZATION | 51.43 | 98.23 | 98.51 |
| PEOPLE | 33.19 | 99.61 | 99.70 |
| TOPICS | 81.80 | 99.19 | 99.29 |
| PLACES | 72.23 | 99.19 | 99.27 |

From above results, we can say that KNN based learning technique is more suitable than Naïve Bayes and Term Graph classification technique for the mining of text or documents. The accuracy reported for KNN is much high than Naïve based method as shown in Table 1 for each category of the dataset.

## 4. Conclusion

We conclude that KNN shows the maximum accuracy as compared to the Naive Bayes and Term-Graph. The drawback for KNN is that its time complexity is high but gives a better accuracy than others. We implemented Term-Graph with other methods rather than the traditional Term-Graph used with AFOPT. This hybrid shows a better result than the traditional combination. Finally we made an information retrieval application using Vector Space Model to give the result of the query entered by the client by showing the relevant document. We will focus more in future on Reducing Complexity, Increasing Accuracy and Text Summarization.

## Authors

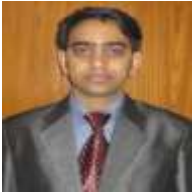

**Vishwanath Bijalwan** is working as assistance professor at Uttarakhand Technical University Dehradun. He is person with lot of potential. His research interest is wireless sensor network, WiMax, Wi-Fi, machine learning, Informtion Retrieval. He obtained B.Tech. and M.Tech from DIT Dehradun. So far he has published 4 high quality researches paper and work on many MHRD funded project. He is carrying 5year of research experience.